\documentclass[a4paper]{spie}  
\usepackage{wasysym} 
\usepackage[]{graphicx}
\usepackage{lscape}
\usepackage{subfig}

\title{Wavelength calibration of the JWST-MIRI medium resolution spectrometer} 

\author{J.~R. Mart\'{\i}nez-Galarza\supit{$\star$a}, A.~M. Glauser\supit{b,c},  A. Hern\'an-Caballero\supit{d}, R. Azzollini\supit{d}, A. Glasse\supit{c}, \\S. Kendrew\supit{a}, B. Brandl\supit{a}, F. Lahuis\supit{e}
\skiplinehalf
\supit{a}Leiden Observatory, Leiden University, P.~O. Box 9513, 2300 CA Leiden, The Netherlands
\skiplinehalf
\supit{b}Institute of Astronomy, ETH Zurich, 8093 Zurich, Switzerland
\skiplinehalf
\supit{c}UK Astronomy Technology Centre, Blackford Hill, Edinburgh EH9 3HJ, United Kingdom
\skiplinehalf
\supit{d}Consejo Superior de Investigaciones Cient\'{\i}ficas, Serrano, 121 28006 Madrid, Spain
\skiplinehalf
\supit{e}SRON Netherlands Institute for Space Research, P.~O. Box 800 9700 AV Groningen, The Netherlands
}

\authorinfo{*martinez@strw.leidenuniv.nl}

\begin{document} 
 \maketitle 

\begin{abstract}
We present the wavelength and spectral resolution characterisation of the Integral Field Unit (IFU) Medium Resolution  Spectrometer for the Mid-InfraRed Instrument (MIRI), to fly onboard the James Webb Space Telescope in 2014. We use data collected using the Verification Model of the instrument and develop an empirical method to calibrate properties such as wavelength range and resolving power in a portion of the spectrometer's full spectral range (5-28$\: \mu$m). We test our results against optical models to verify the system requirements and combine them with  a study of the fringing pattern in the instrument's detector to provide a more accurate calibration. We show that MIRI's IFU spectrometer will be able to produce spectra with a resolving power above $R=2800$ in the wavelength range $6.46-7.70\: \mu$m, and that the unresolved spectral lines are well fitted by a Gaussian profile.
\end{abstract}



\section{INTRODUCTION}
\label{sec:intro}  

The James Webb Space Telescope (JWST) is the next milestone in optical/IR space astronomy. Scheduled for launch in 2014, this NASA/ESA/CSA cryogenic mission will study the infrared (0.6-28$\: \mu$m) universe with unprecedented resolution and sensitivity. Its 6.5$\: $m mirror will provide spatial resolutions of 0.05 arcseconds in the near infrared (0.6-5$\: \mu$m) and 0.11 arcseconds in the mid-infrared (5-28$\: \mu$m), as well as a detection limit of 0.2$\: $mJy at 5.6$\: \mu$m. JWST will carry four instruments onboard: A Near InfraRed Camera (NIRCam), a Near InfraRed Spectrograph (NIRSpec), a Mid InfraRed Instrument (MIRI) and a Fine Guidance Sensor (FGS).

MIRI~\cite{Wright08} will comprise an imager (MIRIM) with a 1.3 x 1.9 arcmin$^2$  field of view, a low-resolution ($R \sim 100$) slit spectrometer (LRS), a medium-resolution ($R \sim 3000$) spectrometer (MRS) using an Integral Field Unit (IFU) and several coronagraphic masks. When functional, MIRI will provide imaging and spectroscopy with unprecedented sensitivity in the mid-infrared portion of the electromagnetic spectrum, with wavelength coverage in the range 5-28$\: \mu$m.  The MRS~\cite{Wells06} is designed to be two orders of magnitude more sensitive than Spitzer Space Telescope's Infrared Spectrograph (IRS), which operated in a similar spectral range, and to have a factor of about 4-6 higher spectral resolution. 

MIRI's optics have been designed and built by the MIRI European Consortium, which includes a number of European universities and institutions. A Verification Model (VM) of the instrument was tested for performance and functionality in two extensive campaigns during 2008 and 2009 at the Rutherford Appleton Laboratry (RAL), in the United Kingdom.  The VM is an identical copy of the instrument useful to state the functionality of MIRI. Similar tests will be carried out during the first half of 2011 for the Flight Model (FM) of the instrument, which is currently being assembled at RAL. To simulate the telescope beam, a MIRI Telescope Simulator has been designed and assembled by the Instituto Nacional de T\'ecnica Aeroespacial (INTA), in Spain. Even though many of the performance tests will be repeated when JWST is flying,  early knowledge on some crucial aspects such as wavelength characterisation, sensitivity, straylight, and image latency will be very useful for the definition of the on-flight calibration plan, software development, etc. Of particular interest are the wavelength and spectral resolution calibration of the Medium Resolution Spectrometer (MRS), since the optical models alone only provide us with a partial understanding of the instrument.

In this paper we present VM MRS spectral data and describe a method to determine the wavelength calibration and spectral resolving power using these data in combination with other reduction tools developed by the team. We apply our method to channel 1C (see Table 1) of the MRS, derive wavelengths and compare with the optical models and the instrument requirements. The method seems reliable and able to provide calibration within the required uncertainties as long as enough signal to noise ratio is achieved in the data. During the Flight Model testing, when all channels will be available and full illumination of the MRS field of view will be achieved, our method will be crucial in the determination of a full range wavelength calibration and included in the data reduction pipeline.

The paper is structured as follows. In \S 2 we describe the spectrometer and its setup during VM testing. We also present the resulting measurements. \S 3 describes the method we used to perform wavelength calibration, and its application to channel 1C. We aso discuss fringing, resolving power and line shape. Finally, we summarize our results in \S 4.

\section{SETUP AND MEASUREMENTS} 
\label{sec:setup}

We start this section with a brief description of the Medium Resolution Spectrometer (MRS) operation and the MIRI Telescope Simulator. We only mention the relevant aspects for our discussion. Later on, we describe the exposures we obtained during VM testing to perform the wavelength characterization. For a complete description of the spectrometer optics and mechanisms, we refer the  reader to the MIRI desidn and development paper~\cite{Wright08}. 

\begin{table}[t]\label{subchannels} 
\begin{center} 
\begin{tabular}[width=1in]{c c c c } \hline \hline

 \textbf{Subchannel} &  \textbf{Wavelength Range $\mu$m} &  \textbf{No. of Slices} &  \textbf{FoV arcsec$^2$}\\ \hline

\textbf{1A} &  4.87-5.82 &  21 &  $3.7\times 3.7$ \\
\textbf{1B} &  5.62-6.73 &  21 &  $3.7\times 3.7$ \\
\textbf{1C} &  6.49-7.76 &  21 &  $3.7\times 3.7$ \\
\textbf{2A} &  7.45-8.90 &  17 &  $4.5\times 4.7$ \\
\textbf{2B} &  8.61-10.28 &  17 &  $4.5\times 4.7$ \\
\textbf{2C} &  9.94-11.87 &  17 &  $4.5\times 4.7$ \\
\textbf{3A} &  11.47-13.67 &  16 &  $6.1\times 6.2$ \\
\textbf{3B} &  13.25-15.80 &  16 &  $6.1\times 6.2$ \\
\textbf{3C} &  15.30-18.24 &  16 &  $6.1\times 6.2$ \\
\textbf{4A} &  17.54-21.10 &  12 &  $7.9\times 7.7$ \\
\textbf{4B} &  20.44-24.72 &  12 &  $7.9\times 7.7$ \\
\textbf{4C} &  23.84-28.82 &  12 &  $7.9\times 7.7$ \\ \hline

\end{tabular} \caption{Nominal characteristics of the MRS subchannels} 
\end{center} 
\end{table}

\subsection{The MIRI spectrometer and the Telescope Simulator}

The MRS is designed to obtain spatial and spectral data simultaneously in a small region of the detector array, conveniently located next to the MIRI imager field of view for targeting purposes. It has two identical channels (5-12$\: \mu$m and 12-28$\: \mu$m), each subdivided into 2 channels with dichroics, for a total of 4 spectrometer channels, whose simultaneous fields of view are overlapping on a range from $3.6" \times 3.6" $ to about $7.6" \times 7.6" $. The integral field unit has an image slicer design whose output is collimated and dispersed by first order diffraction gratings. The spectral window of each of the IFU channel is covered by three different gratings, and hence a total of 12 gratings are needed to cover the full spectral range. In Table 1 we list the wavelength subchannels produced by this setup. The spectra are combined onto  two 1k$\times$1k detectors. Figure 1 shows schematically the slicing of the image by the spectrometer optics.

\begin{figure}[t]
\begin{center}

\includegraphics[scale=0.4, angle=270]{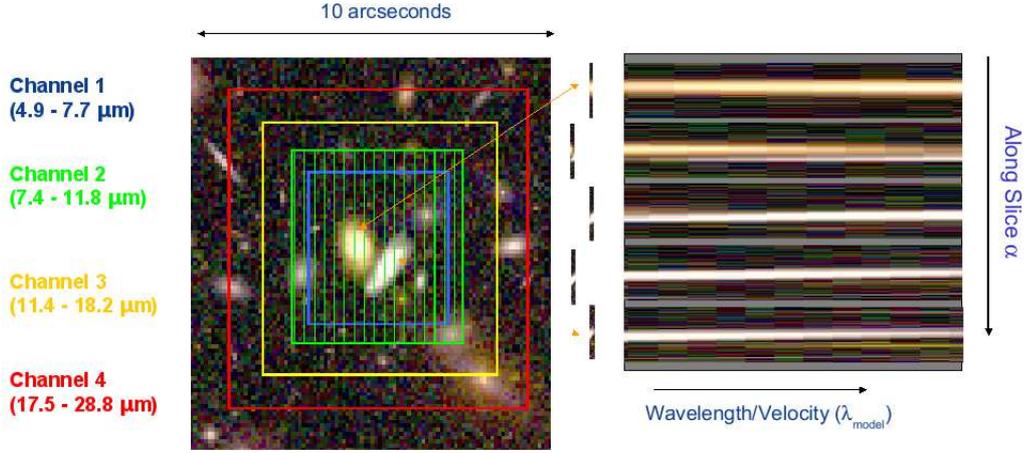}

\caption{\label{IFU_spec}IFU spectrometry. The slicing of the FoV is performed by the spectrometer pre-optics and arranged in an output format, for wach of the 4 concentric FoVs (left). Light is dispersed by first-order gratings and forms a spectrum on the detector (right).} 
\end{center}
\end{figure}

To provide the Verification Model with an input signal, a MIRI Telescope Simulator (MTS)~\cite{Belenguer08} has been developed by INTA. Its main objective is to deliver a test beam to MIRI similar to the one that will be delivered by JWST under flight conditions. Light from a 800$\: $K blackbody is collimated and passed through a variable aperture system and then trough the desired filter before a diffuser spreads the beam and makes the illumination uniform. A 100$\: \mu$m pinhole on the extended source target provides the point-source capabilities. This point source can be scanned across the MIRI FoV. An imaging subsystem then sends the light to the MIRI FoV through a set of focusing and folding mirrors.




MTS was meant to project a point source onto the focal plane of MIRI, as well as an extended uniform illumination. Due to some mechanical issues during the manufacturing of the MTS, it was not possible to properly focus the point source on the focal plane, and hence its Point Spread Function (PSF) was significantly distorted. Additional mechanical issues prevented us to illuminate the full focal plane with the extended source mode. Unfortunately, the non-illuminated area of the focal plane includes the field of view of the four MRS channels. The MTS is also equipped with several narrow-band filters, etalons to provide synthetic spectral lines in the full range, and a mask to produce dark images. It also has two cutoff filters, a long wavelength pass (LWP) and a short wavelength pass (SWP) filter. The MTS etalons are Fabry-Perot interferometers that produce synthetic spectral lines at specific wavelengths by interference of two reflected wave fronts. The pattern of this etalon lines can be resolved by the MRS.




\subsection{Measurements}

Using the setup described in the previous section we have taken MRS spectra of the etalon lines for the different available subchannels. Since full illumination of the MRS field of view was not possible, we used the out-of-focus point source to illuminate two regions of the MRS field of view. Our calibration tool will produce a full calibration once FM testing has provided full illumination for all MRS channels. For each of those etalon exposures we also took a background image by moving the point source out of the image field and taking an exposure of the same duration. Additionally, for each of the two point source positions we took exposures using the MTS LWP filter, and the   800$\: $K blackbody continuum, without any filters. The LWP exposures are crucial to fulfil our goal, since the filter cutoff provides an absolute wavelength reference for the calibration. This cutoff wavelength falls within the range of channel 1C, and hence this is the channel we will be studying here.



 

\begin{figure}[h]
  \centering
  \subfloat[]{\label{fig:etalon_det_spec_et}\includegraphics[width=0.54\textwidth]{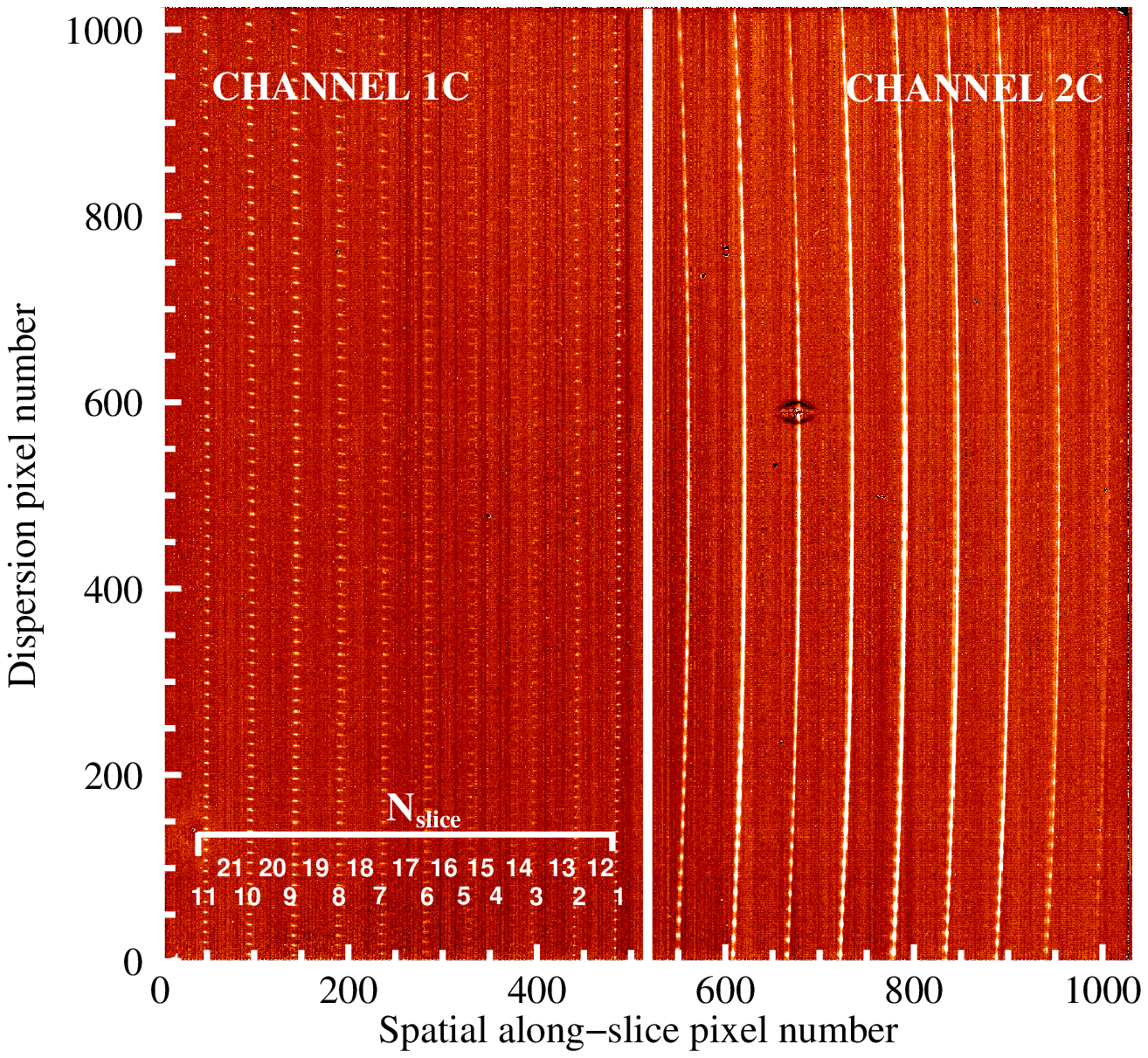}}\subfloat[]{\label{fig:etalon_det_spec_lwp}\includegraphics[width=0.54\textwidth]{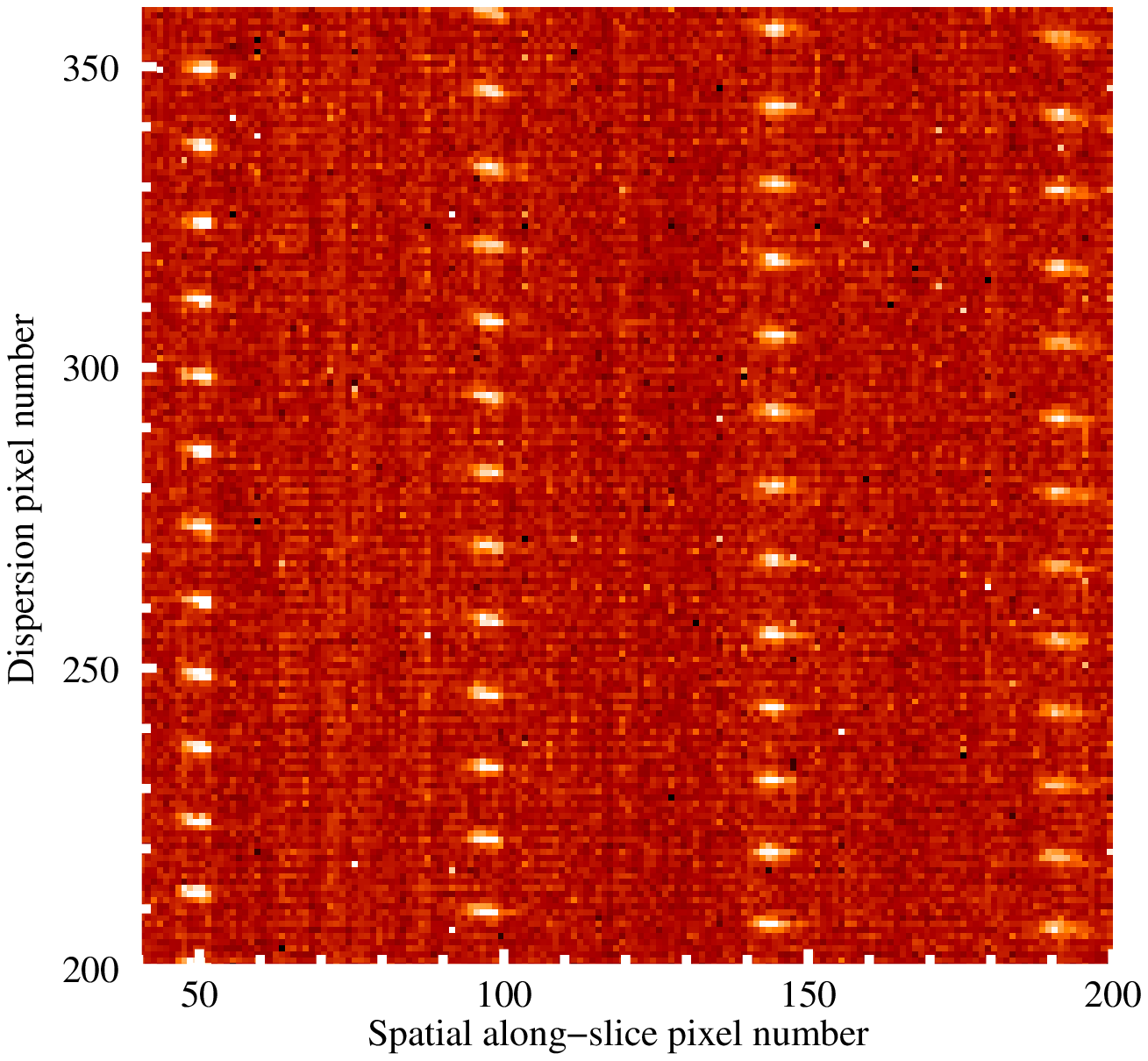}}
  \caption{(a): An etalon exposure using the MRS for channels 1C and 2C. The horizontal axis corresponds to the spatial along-slice direction, while the vertical axis corresponds to the dispersion direction. The slice numbers are indicated (b): Detail of the same etalon exposure. Etalon lines appear as little blobs aligned along the dispersion axis.}
  \label{fig:etalon_det}
\end{figure}

Figure \ref{fig:etalon_det_spec_et} shows the resulting detector signal for channels 1C adn 2C for an etalon exposure. The slices into which the FoV has been divided are arranged along the horizontal axis, while the vertical direction corresponds to the dispersion axis. Each of the vertical 'slices' on the detector plane is then the spectrum of one spatial slice of the FoV. The etalon lines are clearly visible as dots more or less equally spaced along the dispersion axis. They are not present in all the spatial slices, since not all the FoV was illuminated, as mentioned. Optical distortions within the MRS are evident as a curvature of the spectrum. We expect these distortions to produce an observable variation in the position of equal-$\lambda$ points as we move in the along-slice (spatial) directions.

\section{ANALYSIS} 
\label{sec:analysis}
In the first part of this section we describe the method to calibrate the wavelengths for channel 1C, using the etalon and LWP filter exposures. We also study the variations of the wavelength calibration with position in the FoV. In the second part, we use the fringing pattern arising from multiple reflections within the detector layer as an alternative method to calibrate the wavelengths and compare the results with the first method. In the final part of the section we study the shape of the etalon lines obtained with the MRS and discuss the implications for the resolving power of the instrument.

\subsection{Etalon analysis}
From the MRS exposures that we have described in the previous section we can reconstruct the image of the point source on the FoV at different wavelengths. The resulting set of data, a 2D spatial map for each wavelength bin, is known as a datacube. The transformation from pixels on the detector plane (Figure \ref{fig:etalon_det_spec_et}) into the 3D cube space must take into account the optical distortions introduced by the spectrometer pre-optics and main optics, and is obtained via an optical model of the MRS. In (\cite{Glauser10}), the process of how we have performed the mapping to produce a datacube is described and the software tools developed for this purpose are presented. 

We start our analysis with the datacubes produced using these tools from background-subtracted exposures. We use the following coordinate system for the data cube: $\alpha$ is the space coordinate in the along-slice direction, $\beta$ is the space coordinate in the across-slice direction and $\lambda_{\rm{model}}$ is the wavelength. Both $\alpha$ and $\beta$ are measured in arcseconds and are referred to the center of the FoV, while the wavelengths are in microns and their absolute value comes from the optical model. Here we will compare the optical model wavelengths $\lambda_{\rm{model}}$ with the wavelengths resulting from our etalon and LWP measurements and analysis. We refer to the measured wavelengths as $\lambda_{\rm{meas}}$, to differentiate them from the theoretical wavelengths.

The dimensions of the cube are 25 pixels in the $\alpha$ direction, 21 pixels in the $\beta$ direction and 1225 wavelength bins. It is important to note that each of the 21 pixels in the across slice direction ($\beta$) is associated with one of the 21 slices shown in Figure \ref{fig:etalon_det_spec_et}. The spatial sampling is 0.18 arcsecs in the $\alpha$ direction and 0.17 arcsecs in the $\beta$ direction. This results in an effective FoV size of $4.48"\times 3.56"$.

\begin{figure}[h]
  \centering
  \subfloat[]{\label{fig:cube_slice_point}\includegraphics[width=0.54\textwidth]{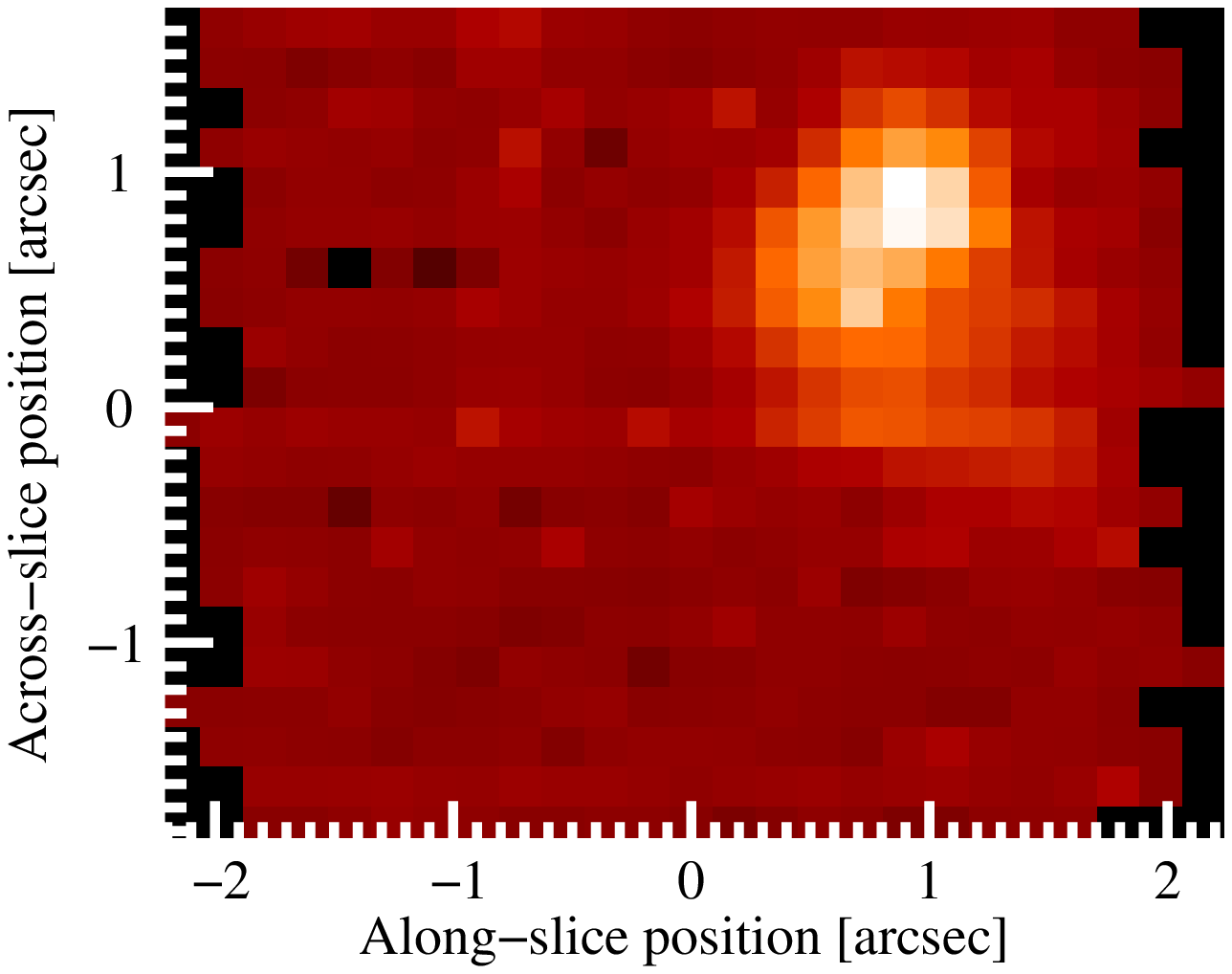}}\subfloat[]{\label{fig:cube_slice_back}\includegraphics[width=0.54\textwidth]{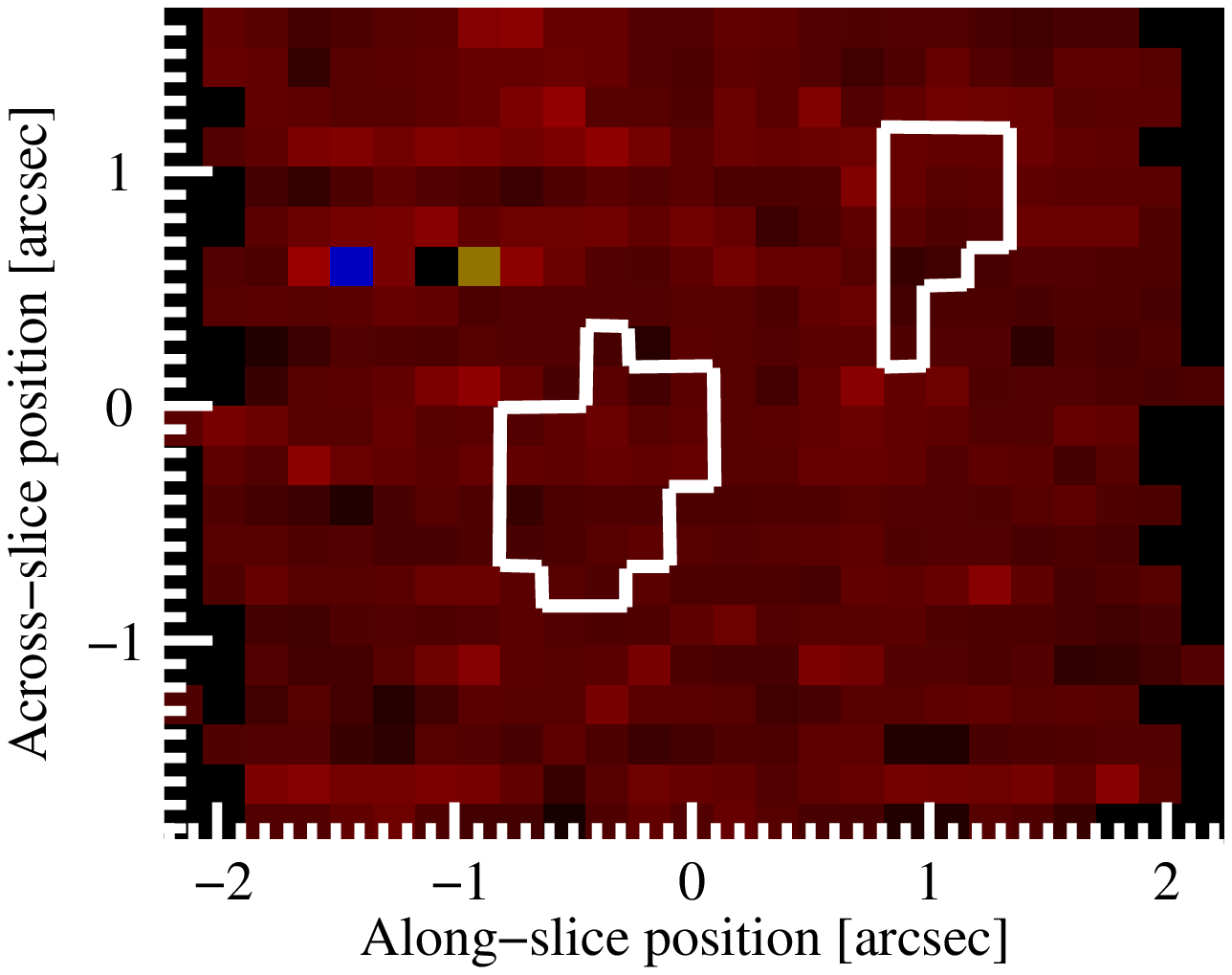}}
  \caption{(a): The reconstructed point source at a particular wavelength. The elongated shape is due to focusing problems of the MTS. The horizontal axis corresponds to the along-slice spatial coordinate, $\alpha$, while the vertical axis corresponds to the across-slice spatial coordinate, $\beta$  (b): The areas of the MRS field of view where enough signal to noise ratio was achieved for this analysis, and where LWP and continuum exposures were also taken.}
  \label{fig:cube_slice}
\end{figure}

Figure \ref{fig:cube_slice_point} shows a layer of the data cube for a particular value of $\lambda_{\rm{model}}$. The original exposure corresponds to a continuum 800$\: $K blackbody exposure. The resulting image shows the point source near one of the corners of the FoV, with an elongated PSF due to the focusing problem of the MTS that we have mentioned before. Figure \ref{fig:cube_slice_back} shows a background exposure where we have indicated the pixels where we have measured the etalon lines with enough signal to noise to perform this analysis, and where we also have a good measurement of the continuum and filter spectra. These areas of the FoV correspond to two different pointings of the point source, and are intended to cover as many pixels as possible.  We will use these well separated regions to study the variation of the wavelength properties and resolving power with position of the point source on the FoV.

The method works as follows: for each available $(\alpha, \beta)$ position we extract the spectrum of the etalon lines exposure and that of the LWP filter exposure. Provided that we know the wavelengths of the etalon lines, they provide a relative wavelength calibration along the full wavelength axis for channel 1C. On the other hand, the cutoff wavelength of the LWP spectrum provides an absolute wavelength reference for a single point in the wavelength axis. We combine the information from this absolute reference with the relative information from the etalon lines to obtain a final wavelength scale. The uncertainty of our method depends on the accuracy with which we are able to identify the etalon line peak positions and the shape of the filter cutoff slope. This accuracy is of course a function of the signal-to-noise (S/N) ratio at which the spectra are detected. Other effects, such as line undersampling, also have to be taken into account.

\begin{figure}[h]
  \centering
  \subfloat[]{\label{fig:spec_extracted_eta}\includegraphics[width=0.375\textwidth, angle=90]{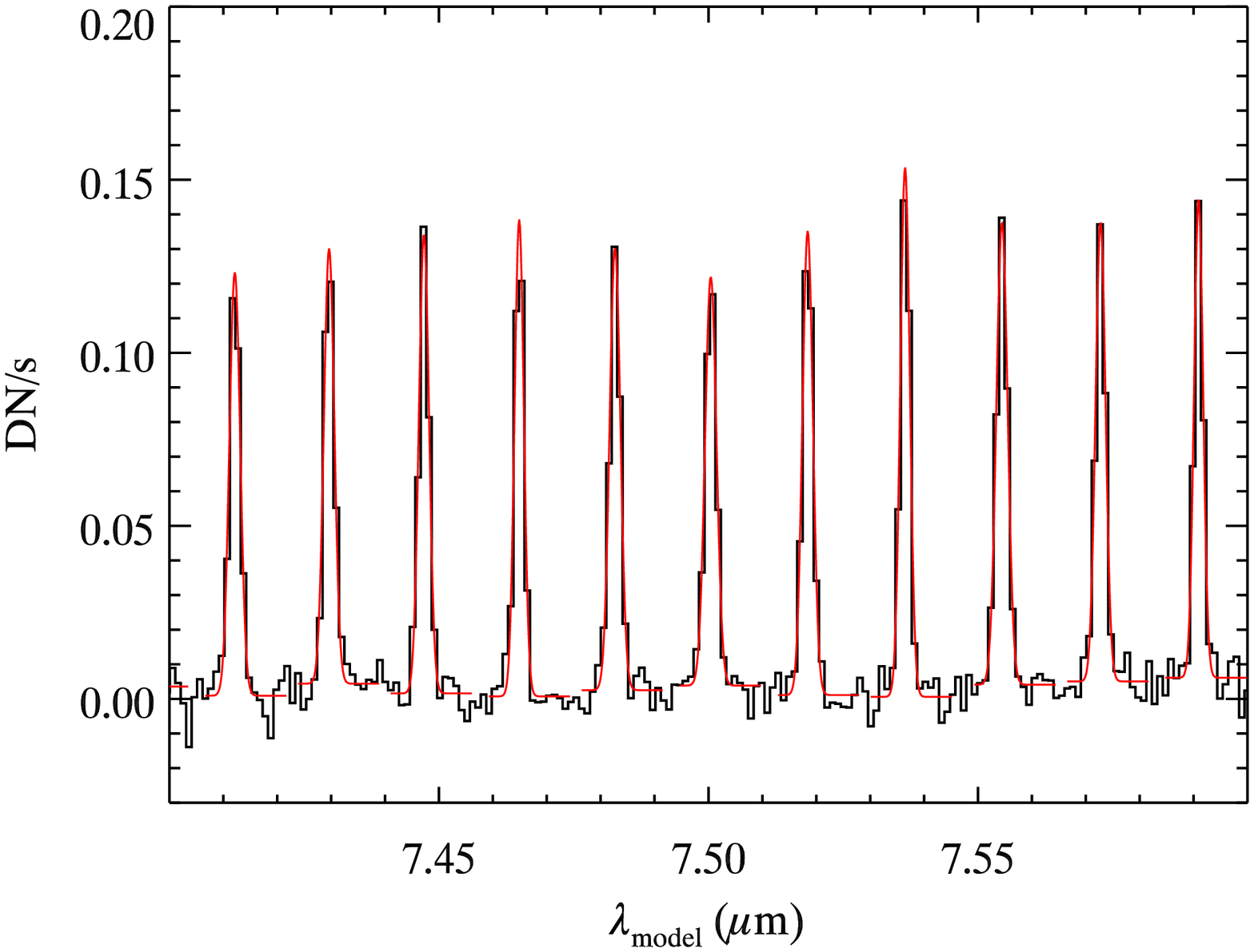}}\subfloat[]{\label{fig:spec_extracted_lwp}\includegraphics[width=0.375\textwidth, angle=90]{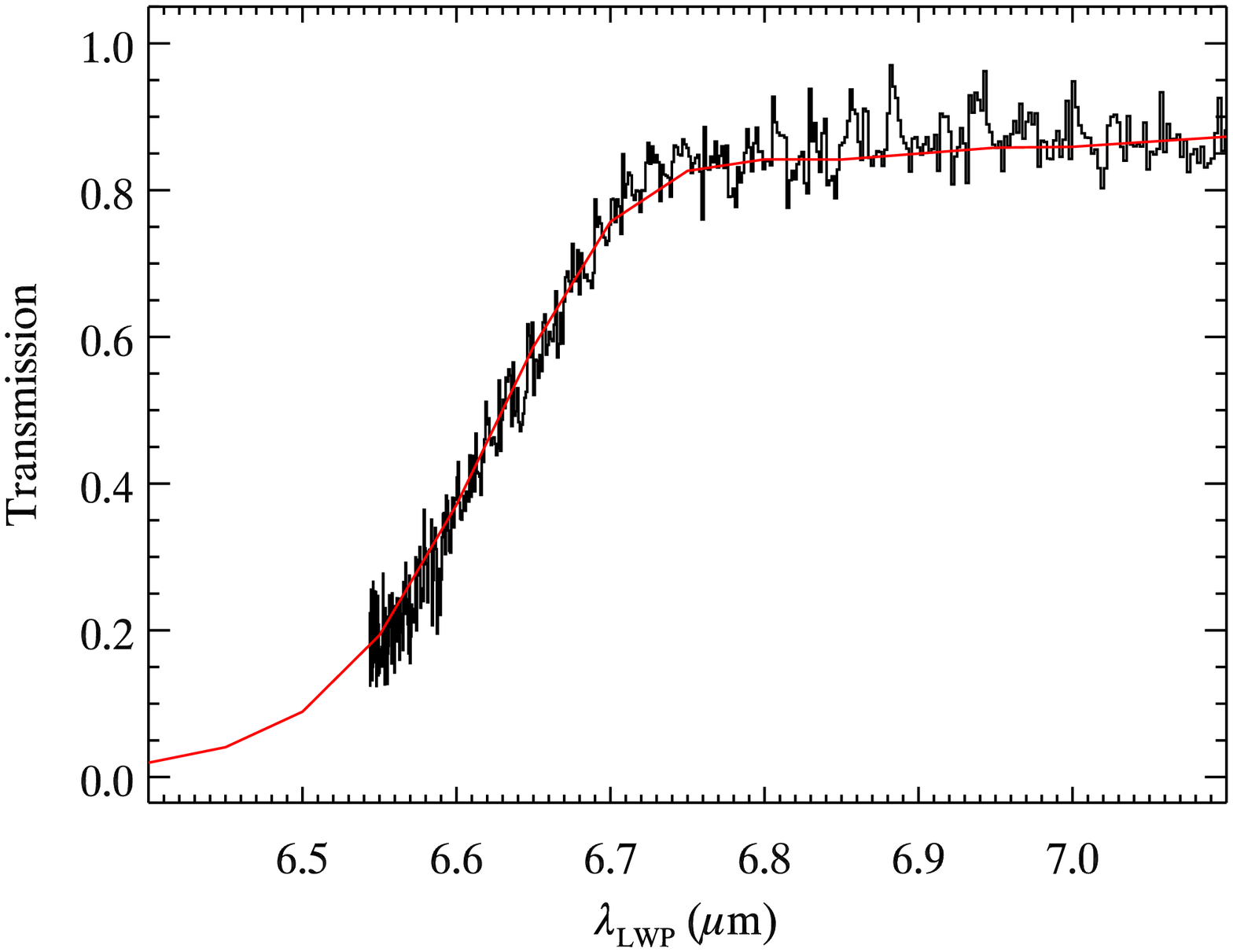}}
  \caption{(a): The extracted etalon spectrum for a particular $(\alpha, \beta)$ value. Signal-to-noise ratio for the etalon lines is about 15. Red lines are Gaussian fits to the data (in black) (b): The transmission profile of the LWP filter. The red line is the calibrated curve, while the black line is the measured data. We fit the two curves to find an absolute reference in wavelength.}
  \label{fig:spec_extracted}
\end{figure}

The left panel of Figure \ref{fig:spec_extracted} shows a segment of the extracted etalon spectrum for $(\alpha=-0.269$ and $\beta=-0.170)$. Lines are well detected with a S/N ratio of about 15, and the separation between them is well resolved, as expected for the MRS resolving power. The pixel sampling of the etalon line profiles adds an uncertainty to our determination of line peaks. Therefore, we fit Gaussian profiles to the lines in order to obtain a more accurate estimate of the peak position. We will discuss the shape of the line profile later in this paper. 

The error in the line position associated with the Gaussian fit is of the order of 0.02 resolution elements. This is, as we will see later, below the instrument requirement for wavelength accuracy. However, the fact that the lines are undersampled and the presence of other sources of noise such as electronics, cosmic rays or bad pixels make the absolute determination of the line positions somewhat more uncertain than the error derived from the Gaussian fit. We expect these lines to be unresolved, and hence, while their centroid positions will provide us with a way to calibrate the wavelengths, their width will give us information about the resolving power of the instrument.

The peak positions of the etalon lines must be related to wavelengths obtained independently with high accuracy. For this purpose, we use a high resolution ($R\sim 10^5$) spectrum of the etalon transmission pattern measured independently at the Rutherford Appleton Laboratory, in the United Kingdom. This measurement provides the relative wavelengths of the etalon lines with an accuracy of the order of 10 angstroms, i.e., far beyond our instrument requirement. Now we only need an absolute reference, and for that we use the cutoff slope of the LWP filter spectrum.

The right panel of Figure \ref{fig:spec_extracted} shows the LWP fit for the position ($\alpha$, $\beta$)$=(-0.2689,-0.1698)$. The cutoff is clearly visible as a fading slope of the flux between 6.5 and 6.7$\: \mu$m. This spectrum is the product of the continuum emission from the 800$\: $K source and the transmission curve of the filter. To obtain an uncalibrated transmission curve from our measurements, we divide the LWP spectrum obtained with the MRS by the 800$\: $K continuum spectrum at the same ($\alpha$, $\beta$) position. We fit the resulting curve to the previously measured filter profile that samples the transmission as a function of the calibrated wavelength $\lambda_{\rm{LWP}}$. It is important to note that this fit does not represent the final dispersion relation yet, but only a first order approximation to find the wavelength of a single bin. Higher orders are not so relevant at this stage, since the accuracy of the determination of the reference bin has to be lower than the separation between etalon lines only (about 0.014$\: \mu$m at 6.6$\: \mu$m).

In the final step we use the position of the etalon line centroids and the wavelength of the reference bin to create a grid of bin number vs. wavelength along the full range of channel 1C. We fit a second order dispersion relation to this grid and apply it to the full set of bins for each available $(\alpha, \beta)$ position. In this way we obtain a calibrated set of wavelengths for the cube bins which are illuminated. This wavelengths ($\lambda_{\rm{meas}}$) are independent from the theoretical wavelengths $\lambda_{\rm{model}}$.

\begin{figure}[t]
\begin{center}

\includegraphics[scale=0.4, angle=90]{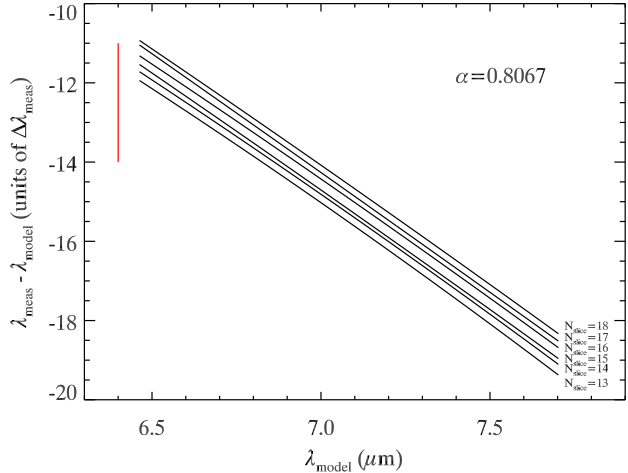}

\caption{\label{diff_slices} Difference between modelled and measured wavelength for different slice numbers, for a particular along-slice position. The length of the red line is half the separation between etalon lines.} 
\end{center}
\end{figure}

Figure \ref{diff_slices} shows the difference between $\lambda_{\rm{model}}$ and $\lambda_{\rm{meas}}$ for different slice numbers, in units of the measured resolution element (see section 3.3). This difference increases linearly as we go to longer wavelengths, and there is a noticeable shift in the difference as we move from one slice to another. The vertical red line in the plot represents half the separation between etalon lines, and hence the minimum error we should expect from effects of line misidentification. The slice to slice variations we are observing are smaller than this minimum error, and hence must arise from different causes. The $\lambda_{\rm{meas}}-\lambda_{\rm{model}}$ difference spans a range of approximately 8 resolution elements for a given slice number, while the offset difference from one slice to another is about 0.2 resolution elements. 


The difference plotted in Figure \ref{diff_slices} is a measure the deviations between the designed and the built spectrometer. For instance, the zero order effect is due to alignment offsets of the detectors or the gratings. The first order effect could be partially explained by a small tilt in the detector, while the second order effects are the result of differences in the optical distortion between the as-build and the as-designed instrument. Obviously, the uncertainty in the knowledge of the LWP cutoff also adds to this difference. The first order effect seems to dominate the difference in this wavelength range.

\subsection{Fringing}

The modulation in intensity along the dispersion axis known as fringing is a common feature in infrared spectrometers and is due to the interference of wavefronts reflected by different layers of the detector. Several de-fringing algorithms have been developed for infrared detectors of different characteristics. We can use the information contained in the fringing pattern in one of  the following ways: either we use the wavelengths of the maxima and minima as compared to the predictions from the optical model to derive accurate detector features such as thickness, refractive index, etc., or we assume a model of the detector and use the distortions in the phase and period of the fringes to obtain a wavelength calibration of the instrument. Here we will attempt this latter option.

We assume that the detector has perfectly plane-parallel surfaces. Under this assumption, the period of the fringes should be constant in the wavenumber ($k$) domain, and deviations from this condition can be used to correct the wavelength.  More specifically, if k is the \textit{correct} wavenumber for each wavelength bin in the cube, then the frequency of the fringes is constant in $k$, and we should observe a linear increase of the phase $\phi$ with $k$, with proportionality factor $\omega$:

\begin{equation}
\phi-\phi_0=\omega(k-k_0)
\end{equation}

We can use this relation to find the wavelength calibration for a particular spatial location in the detector:

\begin{equation}
k= k_0+\frac{\phi-\phi_0}{\omega}
\end{equation}

provided that $k_0$ and $\omega$ are known. From the assumed detector thickness we can obtain the separation between fringes ($k-k_0$), which corresponds to a phase difference of $2\pi$, and hence we obtain $\omega$. As for the absolute reference wavelength $k_0$, we find it again using the LWP filter cutoff wavelength. However, unlike the etalon analysis that uses relative differences between etalon lines, in this case the absolute calibration relies completely on the fit to the cutoff slope of the filter.

\begin{figure}[h]
\begin{center}

\includegraphics[scale=0.7]{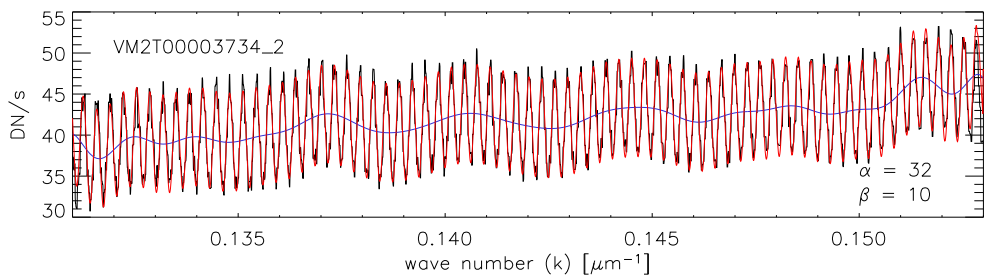}

\caption{\label{fringes} Measurement of the phase of the fringes. The black line represents the counts in the datacube as a function of wavenumber for a given $\alpha$, $\beta$. The blue line is the low-frecuency oscilation obtained through FFT filtering, and the red line is the cosine of the phase calculated for each $k$, $\cos(\phi)$, after scaling to match the amplitude of the fringes and addition of the low-frequency oscilation. Here $\alpha$ and $\beta$ refer to the spatial bin number and the slice number, respectively.} 
\end{center}
\end{figure}

The fringes are obtained from VM exposures using the MIRI calibration source, which provides uniform illumination over the full FoV. Figure \ref{fringes} shows the resulting fringes for a particular value of $\alpha$ and $\beta$. A sinusoidal function is fitted to the baseline-corrected pattern and the position of the fringes maxima is compared with the predictions from the detector model, both in frequency and phase. Deviations are translated into wavelength difference with $\lambda_{\rm{model}}$, as described. 

\begin{figure}[h]
  \centering
  \subfloat[]{\label{fig:fringe_wave_anto}\includegraphics[width=0.375\textwidth]{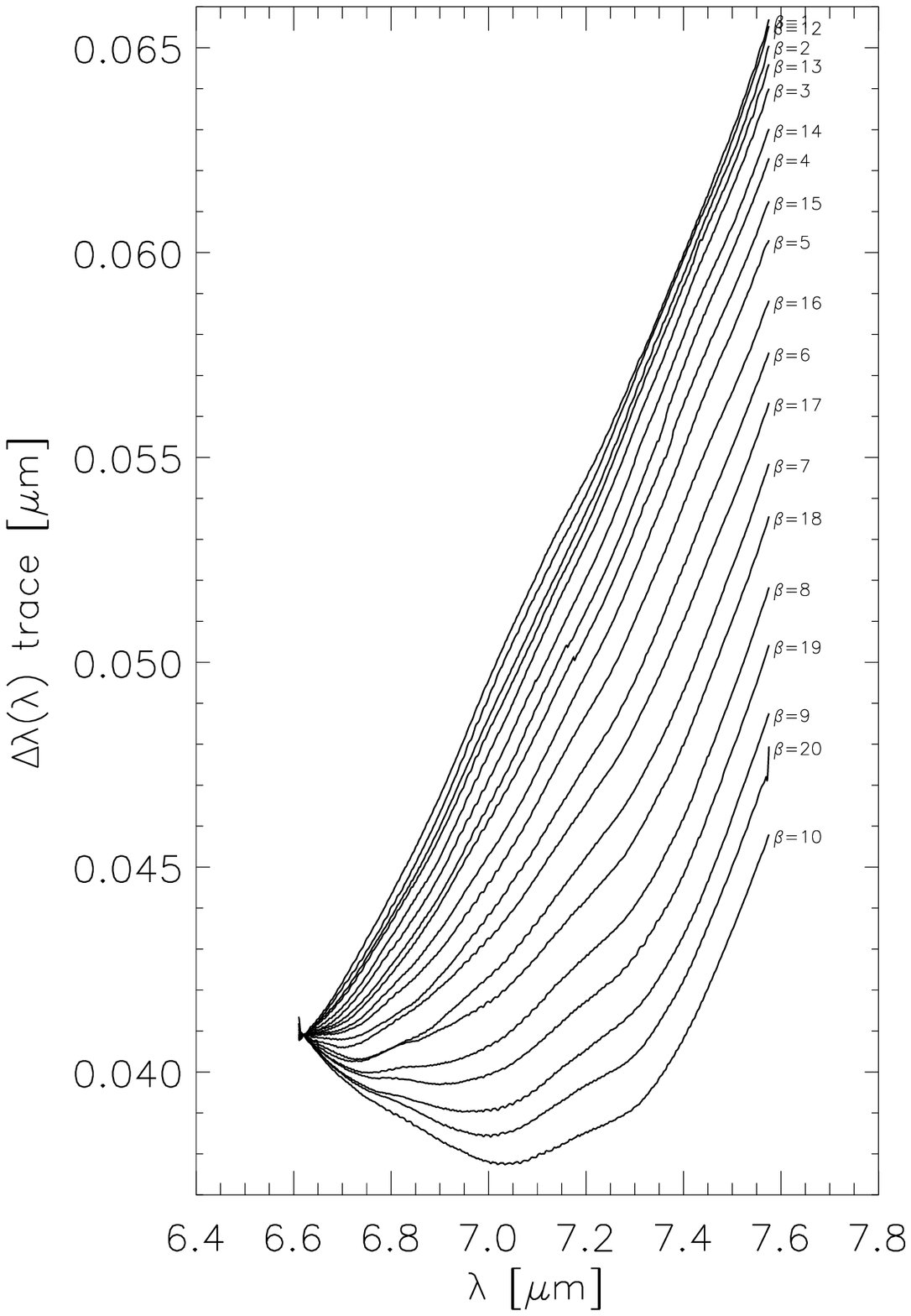}}\subfloat[]{\label{fig:fringe_wave_comp}\includegraphics[width=0.375\textwidth, angle=90]{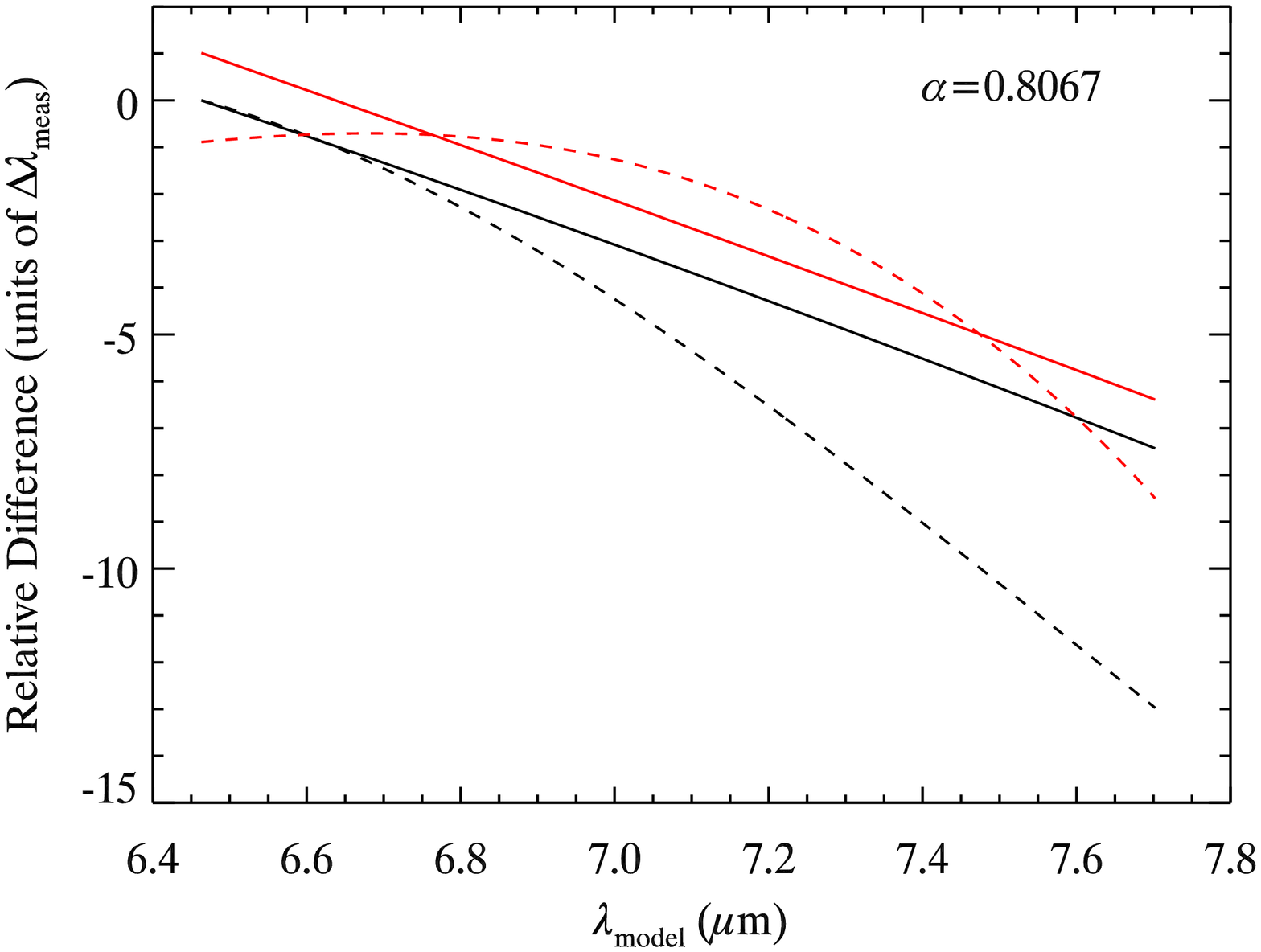}}
  \caption{(a): Shift with respect to $\lambda_{\rm{model}}$ that needs to be applied to each slice as calculated using the fringing analysis. Here, $\beta$ refers to the slice number  (b): Relative difference in the corrections to the model wavelengths when using the two methods described. Solid lines correspond to the etalon analysis, while dashed lines correspond to the fringing analysis. We compare two positions on the MRS FoV: $N_{\rm{slice}}=13$ (black lines), and $N_{\rm{slice}}=18$ (red lines). The origin of the \textit{y} axis is arbitrary.}
  \label{fig:fringe_wave}
\end{figure}

Figure \ref{fig:fringe_wave_anto} shows the correction in wavelengths with respect to $\lambda_{\rm{model}}$ derived using the fringes for all the 21 slices. The shape of the shift varies smoothly as a function of the slice number (i.e., as a function of the slice position on the detector plane). The U-shape of the shift for some of the slices might be due to variations in the detector thickness that are not accounted for in the detector model. But they could also be due to optical distortions in the MRS itself. This result needs to be corroborated during the etalon lines calibration, once the full FoV can be illuminated with the MTS extended source.

We compare the results obtained with the two methods for two different slices in Figure \ref{fig:fringe_wave_comp}. This comparison is only relative, since the strong dependence of the fringing method on the absolute reference makes any absolute comparison meaningless. The difference is significant. The dependence with slice position is much more dramatic when the fringes are used. Also, second order effects seem to dominate the fringing analysis, while the difference with $\lambda_{\rm{model}}$ are close to linear with wavelength in the etalon analysis. The smooth variation in the shape of the curves shown in Figure \ref{fig:fringe_wave_anto} might be an indication that the difference between the two methods might arise from wrong assumptions about the detector properties. For instance, thickness variations (deviations from plane-parallel geometry) of the detector would produce additional distortions on the phase and frequency of the fringes. Whether this is the case, or there are other issues about the wavelength calibration that we are ignoring in the etalon analysis is a question that will have to wait until we have fully characterized FM detectors and better S/N data.

\subsection{Line Shape and Resolving Power}
The etalon lines are not resolved by the MRS. This is shown in Figure \ref{fig:unresolved_gauss}, where we have plotted the high resolution measurement of one of the etalon lines together with the corresponding line as measured with the MRS during the tests. The width of the unresolved lines provides information about the resolving power of the instrument. We have determined the widths of the measured etalon lines by fitting Gaussians to the line profiles and using their full width at half maximum (FWHM). To associate the FWHM of the lines (in pixels) to the resolution element $\Delta \lambda$ (in $\mu$m), we use the derivative of the dispersion relation that we have obtained from the etalon analysis. We compute the resolution element at the positions of the lines and plot the result for a particular ($\alpha, \beta$) in Figure \ref{fig:unresolved_range}, where we have also fitted a strwight line to the set of datapoints. The linear fit gives a resolving power ($R=\lambda/\Delta \lambda$) between 2800 (at short wavelengths) and 3400 (at long wavelengths) for the derived wavelength range (6.43-7.66$\: \mu$m). Deviations in $R$ from the straight line are of about 500. The instrument requirement for resolving power states that between 5 and 10$\: \mu$m, the resolving power should be greater than 2400. The goal, however, is to provide $R\sim3000$ at these wavelengths.

\begin{figure}[h]
  \centering
  \subfloat[]{\label{fig:unresolved_gauss}\includegraphics[width=0.375\textwidth, angle=90]{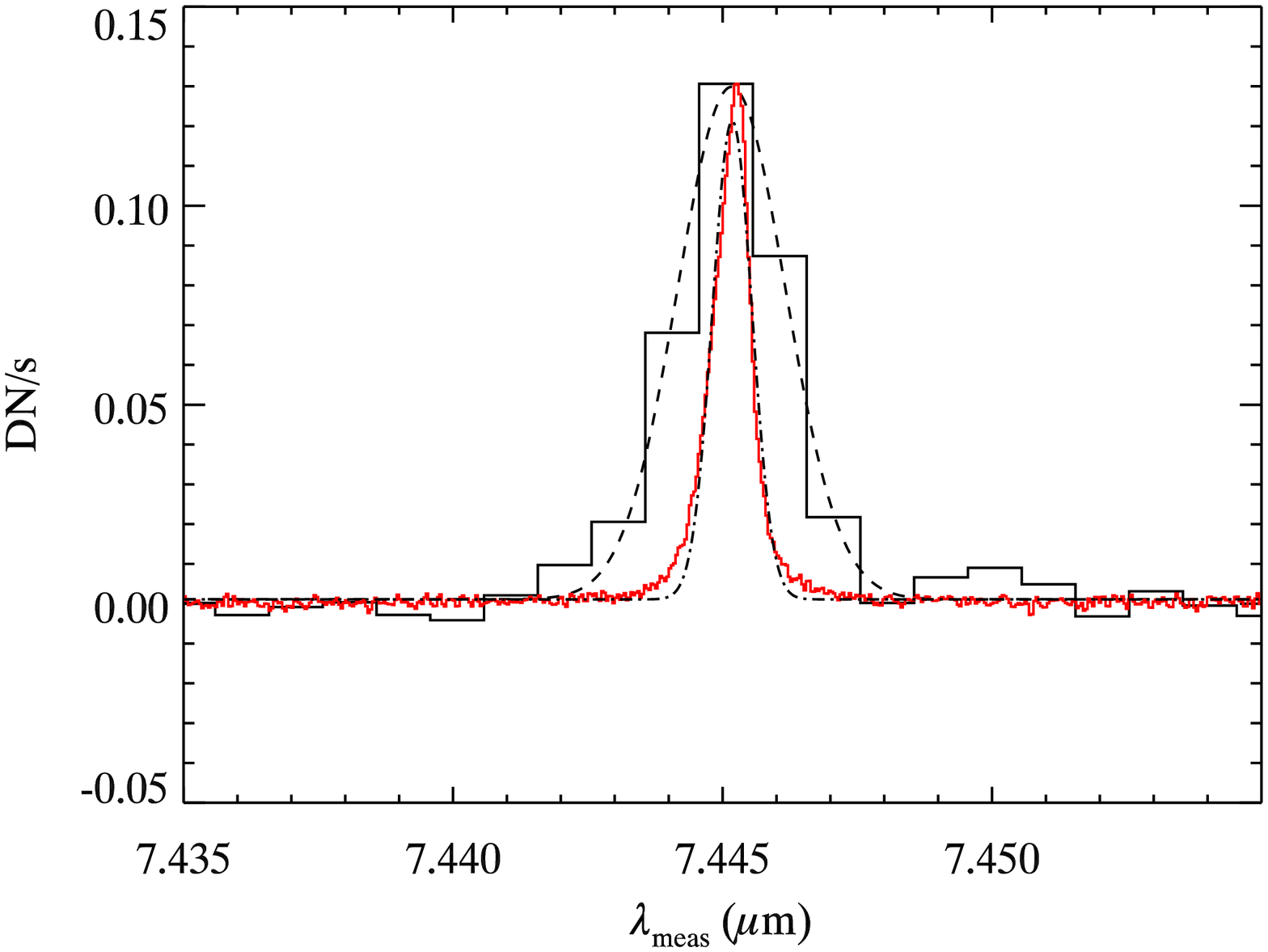}}\subfloat[]{\label{fig:unresolved_range}\includegraphics[width=0.375\textwidth, angle=90]{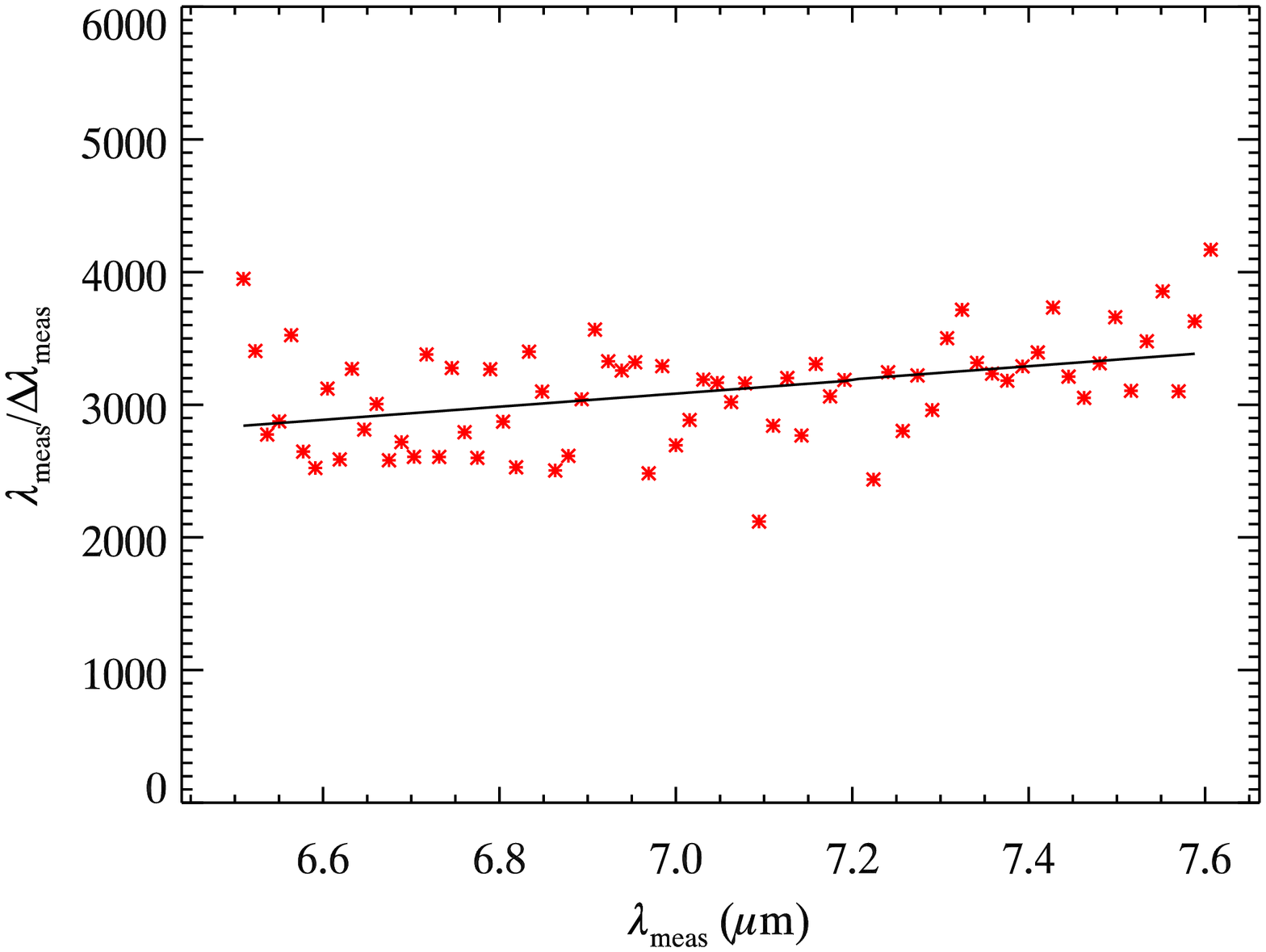}}
  \caption{(a): One etalon line as seen by the MRS (solid black line) and by the high resolution measurements (solid red line). The dotted lines are Gaussian fits to the data (b): Resolving power $R$ for the MRS subchannel 1C.}
  \label{fig:unresolved}
\end{figure}

\begin{figure}[b]
  \centering
  \subfloat[]{\label{fig:unresolved_gauss}\includegraphics[width=0.375\textwidth, angle=90]{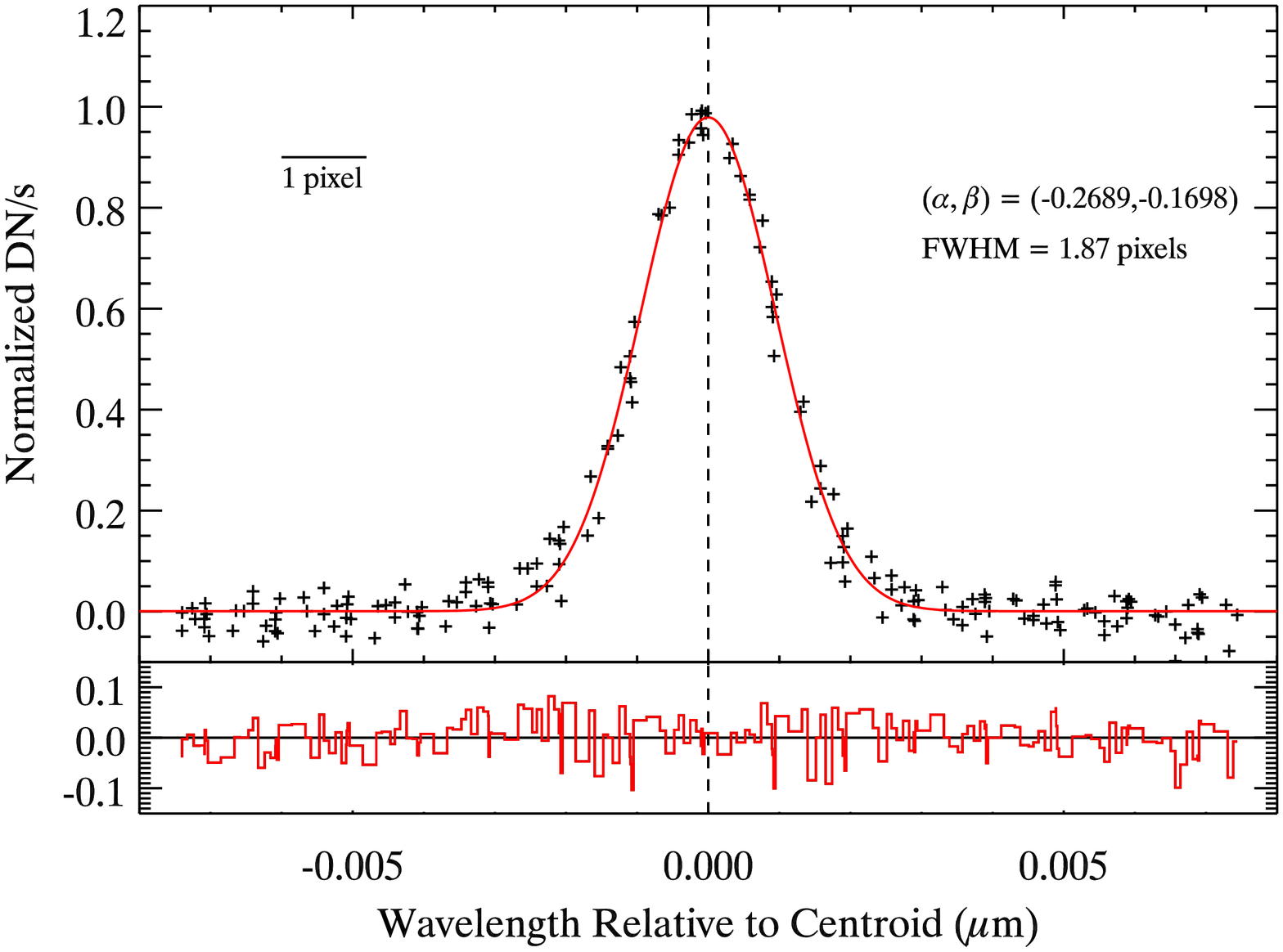}}\subfloat[]{\label{fig:unresolved_range}\includegraphics[width=0.375\textwidth, angle=90]{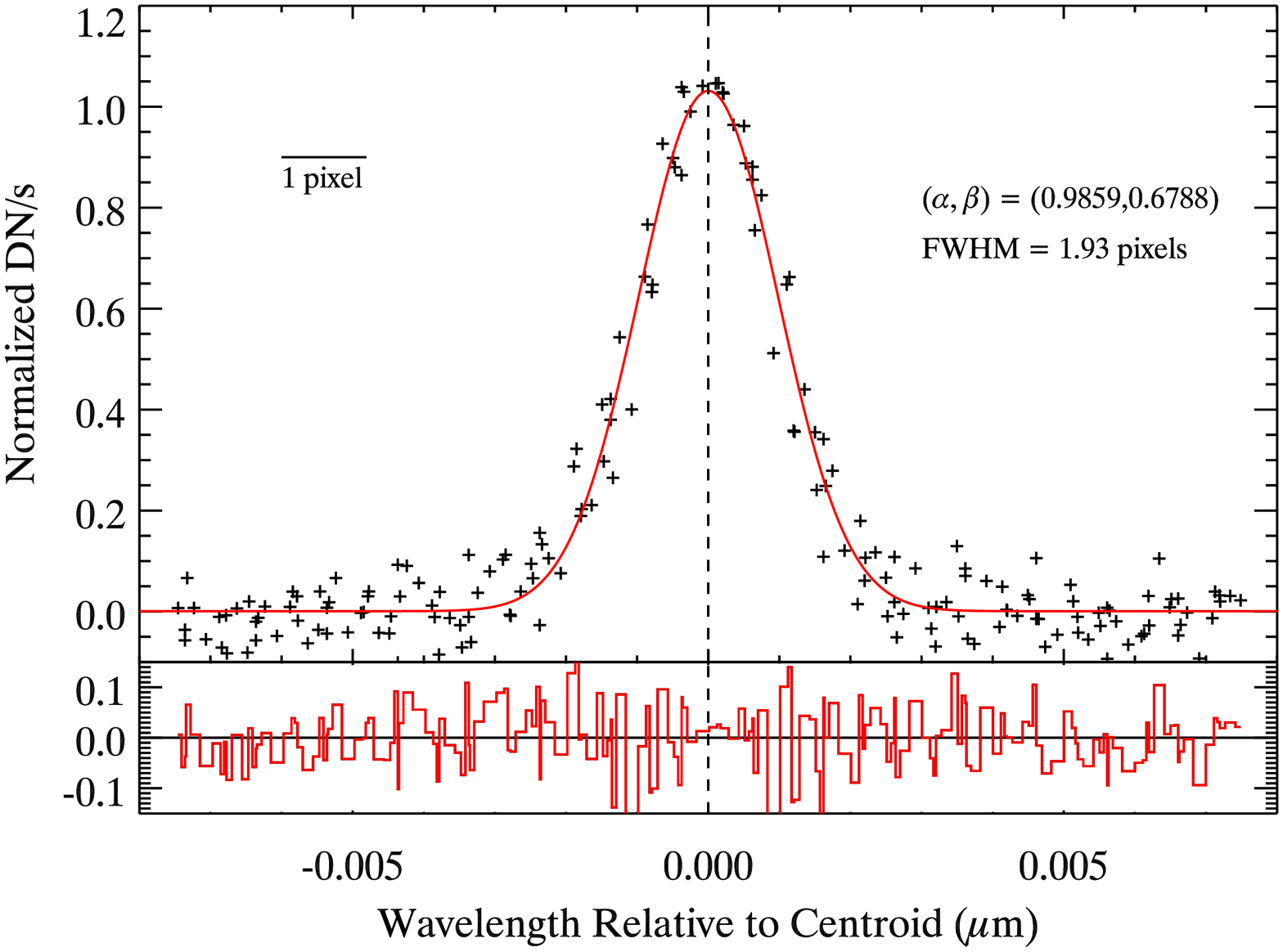}}
  \caption{(a): Overplotted etalon lines with their Gaussian centroids located at the origin, for position ($\alpha$, $\beta$)$=(-0.269,-0.170)$. The cross symbols correspond to the MRS data, while the red line is a Gaussian fit to the resulting set of data. The residuals of the fit are shown at the bottom of the plot. The pixel size is also indicated  (b): Same as (a), but for position ($\alpha$, $\beta$)$=(0.986,0.679)$}
  \label{fig:superres}
\end{figure}

There is a spread in the measurement of the resolving power $R$ of about 500. Insufficient S/N in the measurement of the etalon lines, which causes variations in the parameters of the Gaussian fits from line to line, may be part of the cause for this dispersion. In general, any source of noise in the original exposures can affect the shape of the lines, but we expect the flat-fielding to be the main source of error. Higher S/N again is desirable to achieve a narrower dispersion in the resolving power estimates. During FM testing, studies of the stability of the line shapes is essential to assess what is the magnitude of the variation in resolving power we should expect from one measurement to the other. 

The etalon lines as measured by the MRS are the convolution of the intrinsic line shape with the slit width of the spectrometer, which has a response that we can approximate by a Gaussian. For resolving power issues, we are more interested in getting the instrument's response unaffected by this convolution, and hence we want to deconvolve the measured profile with the intrinsic line shape. The derivation of the intrinsic line shape would require detailed modelling of the reflecting layers of the etalon device mounted on the MTS. Instead, we use the high resolution measurement of the etalon lines as a good approximation of the intrinsic line shape. Figure \ref{fig:unresolved_gauss} shows the Gaussian fits to the unresolved etalon line measured by MRS and the fit to the high resolution measurement for the same line. The theory of convolution of Gaussians states that the widths of two convolved Gaussian add in quadrature: $\sigma_{\rm{convol}}^2=\sigma_1^2+\sigma_2^2$

From the two measurements we obtain with our Gaussian fits and the expression above, we obtain an estimation of the line width, once it has been deconvolved with the intrinsic shape. We find that the deconvolved widths are about 10\% smaller as compared with the MRS measurements. Since the width of the Gaussian is proportional to the resolution element, this decrease in the width translates to an equivalent increase of the resolving power. This increment brings us above our goal of $R>3000$ at these wavelengths. 

\subsubsection{Line shape}
Perhaps more important than an absolute wavelength calibration at this point, the study of the unresolved line shapes across the FoV is crucial. If the line is not symmetric, future dynamical studies using spectral lines might be affected by a purely instrumental effect. For instance, near the edge of the detector, internal reflections might cause additional line peaks or bumps. It is difficult to judge the symmetry of the etalon lines given the level of pixel sampling we have in the data. With FWHMs of between 2 and 3 pixels, the lines might appear assymetric, but they hardly provide a definitive answer. 

In Figure \ref{fig:superres} we have overplotted 12 normalized neighboring etalon lines with wavelengths near 7.4$\: \mu$m with their Gaussian centroids all located at the origin, for two different positions of the FoV. The plots show a statistical line shape that is well fitted by a Gaussian, within the errors shown in at the bottom of the plots.  For both positions, the FWHM of the fitted Gaussias is of about 1.9 pixels.

As mentioned before, undersampling reduces the accuracy with which the line centroids can be determined.  Although geometrical ray tracing predicted linewidths  of less than one pixel, our measurements in Figure \ref{fig:unresolved_range} show that the actual line widths are close to Nyquist sampling, at least in the wavelength range studied here.

\section{Summary and Outlook}
We have studied the wavelength properties of the MIRI Medium Resolution Spectrometer. Using input from optical modelling and an analysis of the synthetic etalon lines produced with the telescope simulator, we have built a tool to obtain a wavelength calibration from MIRI test data. Knowing the wavelength properties of the instrument before its launch is of crucial importance for the development of the reduction software and for the planning of commissioning and science observations that will require the spectroscopic capabilities of JWST.

Even though we were restricted by the quality of the data, which we expect to improve during the testing of the Flight Model of MIRI, the different methods seem to provide the necessary accuracy to verify the instrument requirements. In particular, even at a moderate S/N ratio of about 15, the etalon line analysis provides wavelength references with an accuracy below a tenth of the resolution element. An absolute wavelength calibration will require smaller uncertainties in the determination of an absolute reference using the LWP filter. But even with our S/N restrictions, the effect of a wrong matching of the etalon lines with their actual wavelengths is not bigger than the separation between two etalon lines, and this should be improved with FM data. Fringing provides an interesting way of checking the results, but it might be more useful to study variations in the detector properties such as substrate thickness. The resolution-limited line shapes are well fitted by gaussians and have FWHM larger than one pixel, minimizing the efect of a wrong calibration due to undersampling.

The analysis has been carried out for channel 1C of the detector, where we can establish an absolute reference with the LWP filter cutoff wavelength. However, this method can be extended to other channels by finding an absolute reference for those channels where a filter cutoff is not available. This reference can be, for example, the pattern created by synthetic lines from different etalon filters that overlap in some portions of the different subchannels. In this way, once we have obtained a full set of data with the FM, an absolute wavelength calibration can be achieved for the full wavelength range of MIRI.

\acknowledgments     
 
This work would not have been possible without the input from many people within the MIRI European Consortium and in particular the MIRI test team. Our colleagues in the US have also provided very helpful comments and new ideas. In particular, this work relies strongly on the Data Handling and Analysis Software (DHAS), which among other functions, produces datacubes from detector exposures. This software was developed by Jane Morrison, at the University of Arizona.


\bibliography{report}   
\bibliographystyle{spiebib}   

\end{document}